# Measuring the Density Structure of an Accretion Hot Spot

C. C. Espaillat[1], C. E. Robinson[2], M. M. Romanova[3], T. Thanathibodee[4], J. Wendeborn[1], N. Calvet[4], M. Reynolds[4], & J. Muzerolle[5]

[1] *Institute for Astrophysical Research, Department of Astronomy, Boston University, 725 Commonwealth Avenue, Boston, MA 02215, USA*

[2] *Department of Physics & Astronomy, Amherst College, C025 Science Center 25 East Drive, Amherst, MA 01002, USA*

[3] *Department of Astronomy, Cornell University, Ithaca, NY 14853-6801, USA*

[4] *Department of Astronomy, University of Michigan, 1085 South University Avenue, Ann Arbor, MI 48109, USA*

[5] *Space Telescope Science Institute, 3700 San Martin Drive, Baltimore, MD 21218, USA*

**Magnetospheric accretion models predict that matter from protoplanetary disks accretes onto the star via funnel flows which follow the field lines and shock on the stellar surface[1,2,3] leaving a hot spot with a density gradient[4,5,6]. Previous work has inferred different densities in the hot spot[7], but has not been sensitive to the radial density distribution. Attempts have been made to measure this with X-ray observations[8,9,10], but X-ray emission only traces a fraction of the hot spot[11,12] and also coronal emission[13,14]. Here we report periodic UV and optical light curves of the accreting star GM Aur that display a time lag of about 1 day between their peaks. The periodicity arises as the source of the UV and optical emission moves into and out of view as it rotates along with the star. The time-lag indicates a difference in the spatial distribution of UV and optical brightness over the stellar surface. Within the framework of the magnetospheric accretion model, this indicates a radial density**

**gradient in a hot spot on the stellar surface since different density parts of the hot spot are expected to emit radiation at different wavelengths. These results are the first observational confirmation of the magnetospheric accretion model's prediction of a density gradient in the hot spot and demonstrate the insights gained from focusing on the wavelengths where the bulk of the accretion energy can be observed.**

We conducted a coordinated multi-epoch multi-wavelength observing campaign of the accreting T Tauri star GM Aur, a ~2 Myr old solar analog surrounded by a disk with a large cavity with a radius of ~20 au[15,16,17]. Gas is flowing inside the cavity and eventually reaches the star, as evidenced by moderate accretion rates[18], but most of the solid material in the cavity has grown to mm and larger sizes as indicated by the lack of significant infrared emission[15] that would be present if a substantial amount of smaller dust grains remained. Giant planets are required to open cavities in disks of tens of au in size[19,20], making GM Aur an ideal candidate to study a protoplanetary disk with the properties required to form planets.

Here we present results from our multi-wavelength variability study including *Swift* X-ray and NUV fluxes, *HST* NUV spectra, LCOGT $u´g´r´i´$ photometry, *TESS* photometry, and CHIRON Hα spectra taken contemporaneously over roughly 35 days. We find no significant variation in the X-ray emission (Extended Data Figure 1), ruling it out as responsible for any of the changes seen and so we do not discuss it further. Daily changes appear in the *Swift* NUV, LCOGT $u´g´r´i´$, and *TESS* light curves with a period of about ~6 d (see Methods), consistent with the 6.1 d rotation period of the star[21] and evidence of rotational modulation (Figure 1). The NUV, $u´$, and $g´$ fluxes (red shaded boxes, Figure 1) peak about a day before (see Methods) the *TESS*, $r´$, and $i´$ fluxes (green shaded boxes, Figure 1). Specifically, the UV data peaked on 2019 December 1, 7, 13 (not as well defined due to sparse data), and 19 while the optical data peaked on 2019 December 2, 8, 14 (not as well defined in the *TESS* light curve), and 20. There is also a dip in all of the light curves on Dec 23-25 (gray shaded boxes, Figure 1). This is followed by the disappearance of the UV peaks in the light curves; the optical peaks remain but appear to be lower than those before the dip in the light curves (Figure 1). We put forward that this combination of features in the light curves is due to structure in the hot spot.

The time-lag between the peaks in the UV and optical light curves as well as the eventual disappearance of the UV peaks is the first observational evidence of a density gradient in the hot spot on the stellar surface. UV emission traces the high-density region of the hot spot, and the optical emission traces the low-density region[7]. As the star rotates, we see the different density regions of the hot spot as it goes in and out of view. When the high-density region of the hot spot disappears, the UV emission decreases significantly.

Accretion shock modeling confirms that the high-density region of the hot spot is dominating the UV emission and that the low-density region of the hot spot is dominating the optical emission (see Methods). These models consist of a vertical accretion column close to the surface of the star that is constrained by the stellar magnetic field and divided into three subregions: the pre-shock zone, the post-shock zone, and the heated photosphere below the shock[22]. The top of the accretion column is met by the funnel flow and the base of this accretion column is the hot spot on the stellar surface. There is a peak in both the observed UV emission and the model emission from the high-density region of the hot spot about 1 d before the peak in both the observed optical emission and the model emission from the low-density region (Figure 2). Periodicity in the light curves (see Methods) consistent with the stellar rotation period[21] points to rotational modulation.

This supports our contention that we are seeing a hot spot with a density gradient as it rotates along with the star.

The above is also consistent with MHD simulations which predict that there should be a gradient of density in the accretion flow, column, and hot spot[5,8,23]. We can quantitatively compare the accretion shock modeling results to the size of the hot spot predicted by 3D MHD models[5]. The low density region of the hot spot is predicted to cover 10%-20% of the stellar surface. At high-density, this coverage can drop to less than 1%. Our accretion shock modeling shows that the hot spot surface coverage decreases at larger densities, consistent with the theoretical predictions. Specifically, the low-density region coverage is 10-20% and the high-density region coverage is 0.1% (Figure 2, Extended Data Table 2, see Methods). These accretion hot spot sizes are also roughly consistent with those measured from Zeeman Doppler imaging of young stars[24].

We also qualitatively compare the observed light curves with 3D MHD simulations that have the approximate properties of GM Aur (Figure 3, see Methods). Our observations point to a hot spot with a density gradient where at times the smaller high-density region of the hot spot is no longer visible (it is presumably behind the star) while the larger low-density region is still visible. This is consistent with 3D MHD simulations[5,23,25] which predict there can be a density gradient in the hot spot and that at times parts of the hot spot may not be visible due to stellar rotation (Figure 4, top). The observations are also consistent with simulated light curves (Figure 4, bottom). Like the observations, the simulated light curves are modulated mainly by the stellar rotation and the emission from the high-density region of the spot peaks before the emission from the low-density region. Another piece of evidence of structure in the hot spot is the disappearance of the UV peaks in the light curves after Dec 25 while the optical peaks appear to remain (Figure 1). The position, shape, and structure of the hot spot varies with time, which particularly influences the high-density region of the hot spot, causing it to occasionally disappear (Figure 4, see Methods). Therefore, emission from the low-density region of the hot spot persists while the emission from the high-density region of the hot spot can disappear, as observed here.

The time delay between the peaks in the UV and optical data rules out that the hot spot is azimuthally symmetrical. However, the simulations do not find that the hot spot is systematically asymmetric. Simulations show that the spots may become asymmetric due to complex processes at the disk-magnetosphere boundary (Figure 4, see Methods). The observed systematic behavior may result from a more complex multipole magnetic field closer to the star, and that part of the funnel flow is redirected such that the spots become systematically asymmetric in the azimuthal direction[26,27]. Previous 3D simulations of stars with dipole+octupole fields show if the octupole field dominates near the star, it may redirect the flow and shape the hot spot[28].

Interestingly, there are dips in the *Swift*, LCOGT and/or *TESS* light curves on Dec 21-22 and Dec 23-24 that precede the disappearance of the UV peaks. The first dip on Dec 21-22 is narrow and only seen in the *TESS* data (the *Swift* and LCOGT observations did not overlap in time). We see the start of the second dip on Dec 23 in the *TESS* data and we trace the entire dip in the *Swift* and LCOGT data (dark gray shaded boxes Figure 1). Dust extinction can lead to dips in the light curve[29]. However, GM Aur does not have significant dust in the inner disk[15]. The observed dips are close to the expected minimum flux due to stellar rotation. There may have also been a decrease in accretion and hence hot spot emission around this time leading to a corresponding decrease at the UV and optical wavelengths. This is consistent with the subsequent disappearance of the UV peaks and the somewhat lower optical peaks.

**Figure 1**

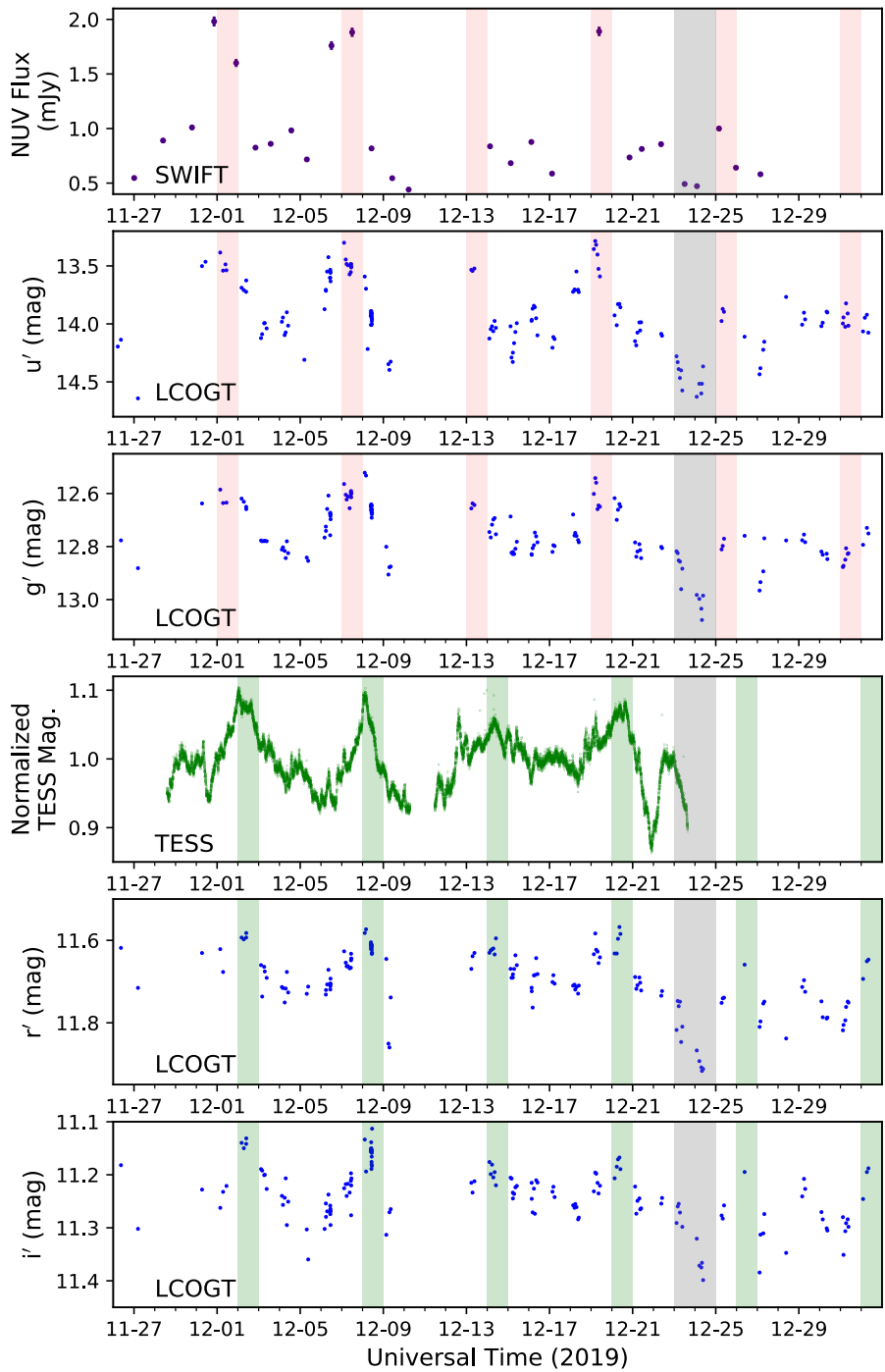

Data from *Swift*, LCOGT, and *TESS* between 2019 Nov 26 to 2020 Jan 01. We show NUV fluxes, *u′ g′* flux calibrated magnitudes, *TESS* normalized magnitudes, and *r′ i′* flux calibrated magnitudes. We note that the *TESS* bandpass covers the bandpass of the *r′* and

*i′* data. Uncertainties for the *Swift* data are listed in Extended Data Table 1; uncertainties are <0.1 mag for the LCOGT and TESS data. Red shaded boxes overlap peaks in the UV light curves and are separated by the rotation period of the star. Green shaded boxes highlight peaks in the optical light curves and are also spaced by the stellar rotation period. Dark gray shaded boxes highlight a contemporaneous dip in all of the light curves.

**Figure 2**

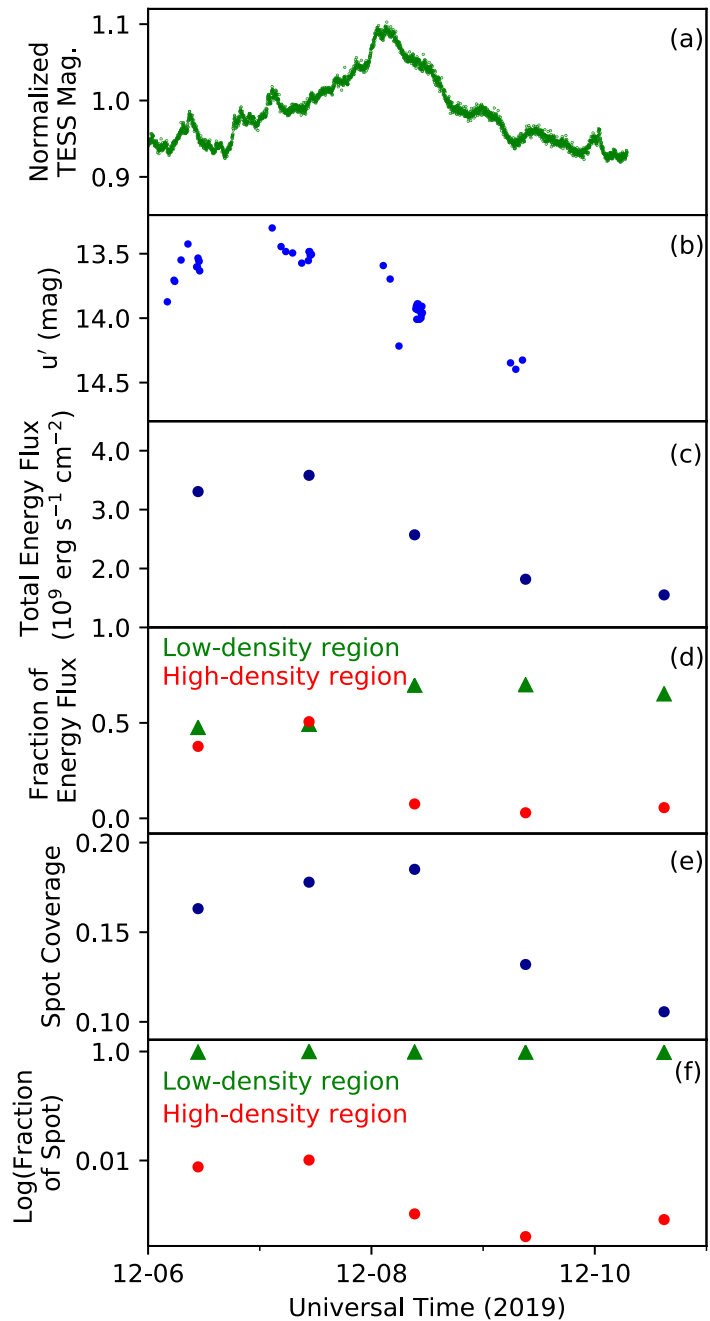

*TESS* optical and LCOGT *u′* data compared to results from accretion-shock-model-fitting (see Methods and Extended Data Table 2) of *HST* spectra over one stellar rotation period.

(a) Optical light curve between 2019 December 6 to 10.

(b) *u′* light curve between 2019 December 6 to 10.

(c) The total energy flux of the accretion column. It peaks along with $u'$.

(d) The fraction of the total energy flux that is due to the low-density region of the accretion column (green triangles) and the high-density region (red circles). The contribution from the high-density region to the total energy flux is highest when $u'$ peaks.

(e) The fraction of the stellar surface covered by the accretion column (i.e., the hot spot). This peaks along with the optical data.

(f) The fraction of the hot spot that corresponds to the low-density region (green triangles) and the high-density region (red circles). The low-density region dominates the hot spot on all days while the high-density region fraction drops significantly after Dec 8.

**Figure 3**

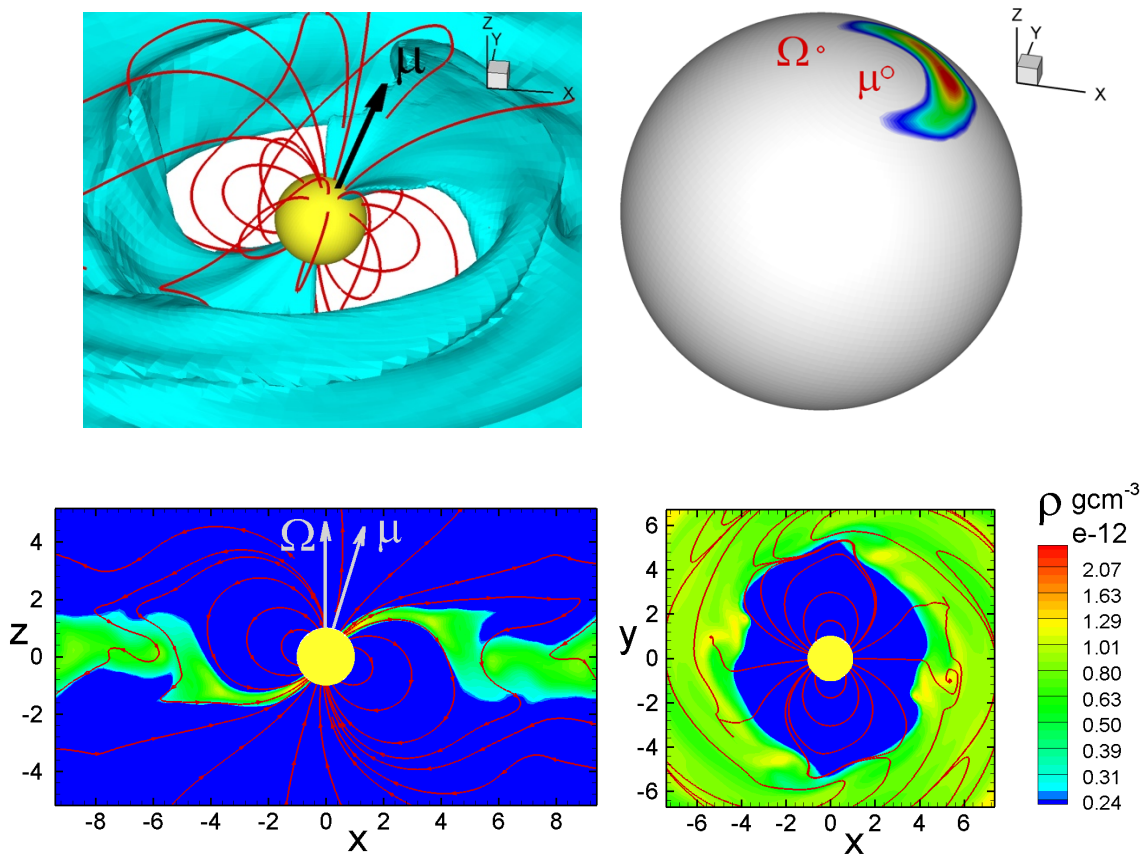

Numerical simulations at a time of 3.13 $P_{star}$. A 3D view (top left) of matter at a density level of $5.3 \times 10^{-13}$ g cm$^{-3}$ (blue) flowing onto the star (yellow) along selected magnetic field lines (red). The corresponding hot spot is also shown (top right) along with slices in the x-z (bottom left) and x-y (bottom right) panels where the color denotes density (see color bar). Here µ is the magnetic moment, Ω is the rotation axis, and ρ is the density. Distance is measured in stellar radii. Only the central part of the simulation region (42 $R_{star}$) is shown for clarity.

**Figure 4**

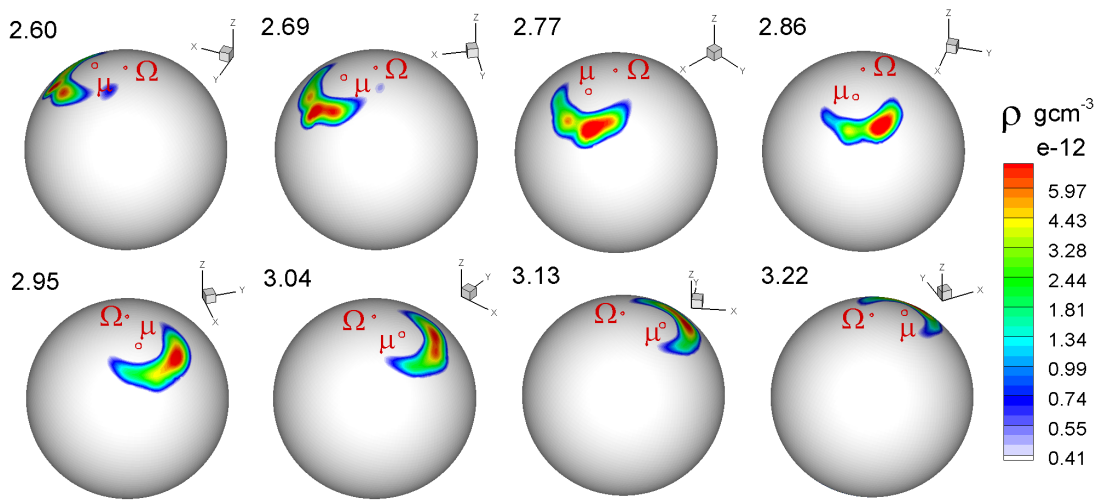

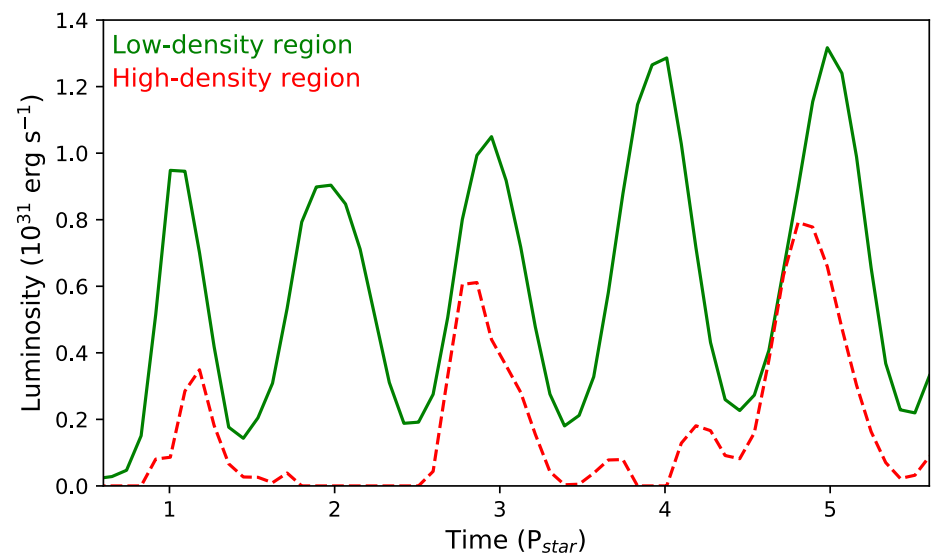

Hot spots (top) and light curves (bottom) from 3D MHD simulations of accretion at the inclination of GM Aur. The hot spot is roughly crescent-shaped and bent around the magnetic pole. There is a density gradient in the hot spot and within a rotation period different parts of the hot spot rotate out of view. Here μ is the magnetic moment, Ω is the rotation axis, and ρ is the density. The light curves generated by the lower density region of the hot spot (green solid) and the densest region of the hot spot (red broken) are mainly modulated by the stellar rotation period. The high-density region of the hot spot can rotate out of view, leading to the offset between the light curves (see ~3 $P_{star}$ and ~5 $P_{star}$).

In addition, the high-density region of the hot spot may occasionally disappear (see ~2 $P_{star}$ and ~4 $P_{star}$, see Methods).

**METHODS**

**Observations and data reduction**

Our campaign of GM Aur was undertaken mostly during 2019 December. Here we present data spanning the X-ray to the optical from *Swift*, *HST*, Las Cumbres Observatory Global Telescope (LCOGT), *TESS*, and SMARTS CHIRON and provide details on the data reduction below.

*Neil Gehrels Swift Observatory* observations were taken with the X-ray Telescope[30] and the Ultraviolet and Optical Telescope[31] using the *UVM2* filter (2221 Å) daily between 2019 November 27 and 2019 December 27 except on days when observations could not be scheduled. For the 27 observations, Observation IDs, exposure times, start times, X-ray count rates, NUV fluxes, and measurement uncertainties are listed in Extended Data Table 1. The Observation IDs are as follows: 0034249045-0034249058, 0034249060-0034249063, 0034249065-0034249070, and 0034249072-0034249074. We utilized the HEASARC HEASOFT software (v. 6.22.1) to measure count rates and NUV fluxes. X-ray emission from the stellar corona is known to be variable[32,33]. X-ray count rates (0.5 – 10.0 keV) of GM Aur usually varied between 0.006 and 0.029 counts $s^{-1}$ (Extended Data Figure 1) and are comparable to those previously seen in GM Aur[34], pointing to an X-ray luminosity of ~$3 \times 10^{30}$ erg $s^{-1}$ (assuming a distance of ~160 pc[35]) throughout the campaign. The count rate was highest on 2019 November 27 (0.046 counts $s^{-1}$). None of the X-ray variability observed appears to correlate with any other data obtained during this campaign (Figure 1, Extended Data Figure 1) and so we do not discuss it further.

*HST* NUV-NIR (1700 – 10000 Å, R~500-1,000) spectra were taken with STIS on 2019 December 6, 7, 8, 9, and 10 (Extended Data Figure 2) in Program 16010. The spectra were obtained with the G230L, G430L, and G750L gratings with a 52''x2'' slit within one orbit per visit. Data were automatically reduced by the *HST* pipeline and here we correct fringing in the G750L data[36]. Archival G230L and G430L spectra of the non-accreting star RECX 1 (Program 11616) was used as the template for the accretion shock modeling analysis below. Typical uncertainties on the *HST* spectra are 3-15%.

*TESS* data were taken on 2019 November 28 to 2019 December 23 (Sector 19) with a cadence of 2 minutes (Figure 1). The *TESS* bandpass covers 6000 – 10,000 Å and is centered at 7865 Å. Data were reduced using the Science Processing Operations Center pipeline. We used Lightkurve[37] to check the data quality and find little contamination from nearby sources. The gap in the data is due to data download during orbital perigee. Typical uncertainties in the *TESS* data are 0.25%.

SDSS $u'g'r'i'$ data were taken roughly 5-10 times on clear nights at the LCOGT[38] with the Sinistro Imagers on the 1 m telescopes between 2019 November 26 and 2020 January 1. The $u'g'r'i'$ filters have central wavelengths of 3540, 4770, 6215 and 7545 Å and wavelength widths of 570, 1500, 1390, 1290 Å. Data in Figure 1 were reduced using the Aperture Photometry Tool and standard aperture photometry techniques. Uncertainties in all bands are <0.1 mag. The flux of GM Aur was calibrated using the fluxes of background objects which in turn were flux calibrated using UBVRI data of background objects (converted to $u'g'r'i'$[39]) and the GD 64 standard field[40]. These calibration data were taken contemporaneously at the 4.3 m Lowell Discovery Telescope (LDT) using the Large Monolithic Imager on 2019 December 2, 7, 10, 13, 18, 21. The LDT data are only used here to flux calibrate the LCOGT data.

Medium-resolution (R~25,000) optical (4082-8906 Å) spectra were obtained with CHIRON[41] on a 1.5m telescope that is part of SMARTS (the Small & Moderate Aperture

Research Telescope System) at Cerro Tololo Inter-American Observatory between 2019 Nov 28 and Dec 17 (Extended Data Figure 3). The standard CHIRON pipeline is not optimized to extract the H$\alpha$ profile of young stars and we reduce the spectra here[42].

**Timing analysis**

While the periodicity and time-lag in the light curves is evident (Figure 1), here we measure the period of the *TESS* and LCOGT $u'g'r'i'$ light curves as well as the time-lag between the peak in the $u'$ light curve and those in the $g'r'i'$ and *TESS* light curves. The *Swift* NUV data is much sparser and we do not analyze it in as much detail.

To measure the period in the light curves, we use the Astropy Lomb-Scargle periodogram function[43,44,45] and obtain a 5.8 d period for the TESS light curve and for the $u'g'r'i'$ light curves we measure periods of 6.3, 6.3, 6.3, and 6.1 d. The *Swift* NUV flux is >50% higher than surrounding days on Nov 30 - Dec 1, Dec 7-8, and Dec 19 (Figure 1) which is consistent with a ~6 day period. Given that the independently measured rotation period of the star is 6.1 d[21] we attribute the ~6 d period measured in the light curves to stellar rotation.

For the time-lag analysis, we use the Python package Stingray[46]. When compared to the $u'$ light curve, we measure time lags of ~0, ~0.75, ~1.1, and ~1 day for the $g'r'i'$ and *TESS* light curves. The *Swift* NUV flux is highest on the same days as the peaks seen in the $u'$ light curve (Figure 1). We conclude that the time-lag between the UV and optical data is ~ 1 d.

**Accretion shock modeling**

To further explore the ~1 d offset in the peak between the UV and optical data seen in Figure 1, we focus on 2019 December 6 to 10 (Figure 2a, b) since *HST* data were taken daily during this time. *HST* NUV spectra are the best measure of accretion since the accretion column that channels material onto the surface of the star emits substantial energy at NUV wavelengths[1].

The *HST* data were fit using accretion shock models[22]. These models provide information on the physical properties of the accretion column and the associated hot spot, which is the footprint of the accretion column on the stellar surface. The accretion column is characterized by an energy flux ($\mathcal{F}=½ \rho v_s^3$) which measures the density of material in the accretion column ($\rho$) assuming that the magnetospheric radius ($R_{mag}$) and infall velocity ($v_s$, here 456 km/s, which depends on $R_{star}$, $M_{star}$, and $R_{mag}$) are constant. Each column has a filling factor, $f$, which gives the fraction of the stellar surface covered by the column (i.e., the hot spot).

GM Aur is a well-studied source in the NUV, with 8 previously modeled epochs[18]. In Figure 2c--f, we plot the accretion-column-model parameters (Extended Data Table 2) obtained from fitting the *HST* data (Extended Data Figure 2). The fitting was done with three accretion columns[7,18] with $\mathcal{F}$ of $1\times10^{10}$, $1\times10^{11}$, and $1\times10^{12}$ erg s$^{-1}$ cm$^{-2}$ and adopting published stellar parameters[47]. Here we refer to the $1\times10^{10}$ and $1\times10^{12}$ erg s$^{-1}$ cm$^{-2}$ components as the "low-density" and "high-density" regions. The total energy flux (Figure 2c) is the sum of the energy flux of all the regions weighted by their respective $f$. The total hot spot coverage on the stellar surface (Figure 2e) is the sum of the $f$ values for all the regions.

The total energy flux peaks along with the UV data (Figure 2c). When the energy flux peaks on December 7, it is dominated by emission from a high-density region, which then drops significantly (Figure 2d). This suggests that on December 7, we see the high-density region of the hot spot and in the following three days, most of this high-density region is no longer visible. Meanwhile, the hot spot coverage of the star peaked ~1 d later on December 8 along with the optical data (Figure 2e), and the hot spot is dominated by emission from a low-density region (Figure 2f). Since the low-density region of the accretion column emits its energy at longer wavelengths[7], it follows that the optical emission peaks when the hot spot is at its largest. The observed behavior in the light curves can be interpreted by combining two effects: the stellar rotation and different physical locations for the high-density and low-density regions. It then follows that the high-density region of the hot spot (whose energy appears mainly in the UV) will lead to a peak in the UV data when the high-density parts of the hot spot are visible to the observer. As the star rotates, we may be seeing a denser part of the hot spot first, which then rotates out of view.

**Accretion flow modeling**

To facilitate comparison to the 3D MHD simulations, we need to estimate the magnetospheric radius of GM Aur. Magnetospheric accretion flow modeling[48,49,50] of the H$\alpha$ line provides properties of the accretion flow. The model assumes that the magnetic, stellar rotation, and disk rotation axes are aligned. The material flows onto the star along an axisymmetric accretion flow arising from the co-rotating gas disk. The geometry of the flow is described by a dipolar magnetic field and characterized by an inner radius ($R_i$, which corresponds to $R_{mag}$) and the width of the flow ($W_r$) at the disk plane. The model assumes a steady flow prescription for a given accretion rate to determine the density at a given point. The temperature at each point is determined parametrically, scaled to the density assuming a constant heating rate in the flow; the maximum temperature in the flow ($T_{max}$) describes each model. To calculate the emission line profile, the model assumes the extended Sobolev approximation and calculates the mean intensity and the level population of a 16-level hydrogen atom and the ray-by-ray method for a given viewing inclination (*i*).

We created a large grid of models varying accretion rate, $R_{mag}$, $W_r$, $T_{max}$, and *i* using the ranges of parameters appropriate for accreting T Tauri stars[49]. With these combinations, we calculated ~72,000 model profiles. We convolved the model profiles with a Gaussian instrumental profile of CHIRON's resolution and fitted each observed profile inside ±400 km s$^{-1}$ from the line center. The best fits are determined by calculating the $\chi^2$ for each combination of the model and observed profile.

For each observed profile, we selected 100 best fits and calculated the mean of the accretion rate, $R_{mag}$, $W_r$, $T_{max}$, and *i* (Extended Data Table 3). The accretion rates listed in Extended Data Table 3 are higher on the dates that there are peaks in the UV emission. The derived *i* are roughly consistent with the measured inclination of the disk which is 53º [17]. $W_r$ varies between 0.2 and 0.5 $R_{star}$ while $T_{max}$ varies between 8270 and 9120 K. We use the derived $R_{mag}$ of GM Aur in order to compare it to simulations in the next section.

In Extended Data Figure 3, we show the best-fitting model to the Hα profiles. The fit to the line wings are very good. The regions with strong absorption features on the blue side of the line were excluded from the fit; this blue-shifted absorption is likely from winds, which are not included in the model. There is no periodic pattern in the Hα line (Extended Data Figure 3). However, there is multi-component high-velocity blue-shifted absorption starting on Dec 7 occurring along with a peak in the UV emission (Figure 1) and accretion rate (Extended Data Table 2 and 3). There is a possible second component at higher, blue-shifted velocity on Dec 6, which might imply a deceleration after the launching of a higher-density outflow event that may presage the accretion event on Dec 7. Models[51] show that there is a disk wind component to the Hα profile which manifests as blue-shifted absorption, and this component is dominated by the innermost disk within tens of stellar radii.

**3D MHD simulations of accretion**

Here we show global 3D MHD simulations of a rotating magnetized star accreting from a disk[5,23,25]. Briefly, the models assume the star has a dipole magnetic field which has a misalignment angle, $\theta$, between the rotation axis, $\Omega$, and magnetic axis, $\mu$. The rotation axis of the star and disk are aligned. Here specifically we use simulations with the same set up as previous work[23,25] with parameters chosen to approximate the properties of GM Aur.

We assume the magnetic field of GM Aur can be approximated with a dipole field. The hot spot of GM Aur is at a high latitude of about 77º, measured from radial velocity variations of He I (5876 Å)[52], and previous work finds that the hot spot is at high latitudes for dipole-octupole configurations[1]. Also, GM Aur has a long rotation period[21] and young stars with simpler, more dipole magnetic fields are slower rotators[53]. Simulations[23] show that the properties of the hot spot depend on (1) $\theta$, (2) the corotation radius, $R_{co}$, and (3) $R_{mag}$. The $\theta$ of GM Aur is 13º [52]. This is consistent with the range of inclinations of the accretion flow inferred from modeling the Hα profile (i.e., all of the best-fit inclinations in Extended Data Table 3 are roughly within ~10º of the system inclination of 53º). $R_{co}$ is 7.8 $R_{star}$ (0.07 au) using a stellar rotation period of 6.1 d[21], $R_{star}$ of 2 $R_{sun}$, and $M_{star}$ of 1.36 $M_{sun}$[47]. We measure a mean $R_{mag}$ of about 3.8 $R_{star}$ with a range of 3.4-4.6 $R_{star}$ (from Extended Table 3). The simulations use θ=20º, $R_{mag}$~4.5 $R_{star}$, and $R_{co}$=5.7 $R_{star}$ which are consistent with the parameters of GM Aur.

The 3D MHD simulations have an approximate accretion rate of ~1.1x10$^{-8}$ $M_{sun}$ yr$^{-1}$, which agrees with measurements from the accretion shock and accretion flow modeling (Extended Tables 2 and 3). The energy distribution in the spots in the 3D MHD simulations varies between ~3.9x10$^9$ to ~1.2x10$^{11}$ erg s$^{-1}$ cm$^{-2}$; this is consistent with the accretion shock models which use energy fluxes of 1x10$^{10}$ to 1x10$^{12}$ erg s$^{-1}$ cm$^{-2}$. The densities of the 3D MHD simulated spots in Figure 4 range between ~5.5x10$^{-13}$ to ~5.97x10$^{-12}$ g cm$^{-3}$ which overlap with the accretion shock models where the densities are 2.1x10$^{-13}$ to 2.1x10$^{-11}$ g cm$^{-3}$. The simulated light curves peak at a luminosity of ~1x10$^{31}$ erg s$^{-1}$ (Figure 4) which is roughly consistent with the accretion luminosities of ~0.1-0.2 $L_{sun}$ (3.8x10$^{32}$ - 7.7x10$^{32}$ erg s$^{-1}$) measured from the *HST* spectra. We note that in Figure 4 we only show the light curve once accretion onto the star begins in the simulation. Also, to generate the light curve for the densest region of the hot spot we use ρ > 6.5x10$^{-12}$ g cm$^{-3}$.

A ratio between $R_{co}$ and $R_{mag}$ of 1.5 sets the boundary between stable and unstable accretion[25]. In the stable regime, matter accretes onto the star in ordered funnel streams and symmetric crescent-shaped hot spots are expected[54]. In the unstable regime, matter accretes in chaotic hot spots[25]. The simulations have $R_{co}$/$R_{mag}$ of 1.3-1.4. The measured $R_{mag}$ of GM Aur has an uncertainty of about 1 $R_{star}$ and so the $R_{co}$/$R_{mag}$ of GM Aur could reach 1.4. Therefore, the simulations are consistent with the properties of GM Aur, and both are near the boundary of the stable/unstable regime. Here, processes at the disk-magnetosphere boundary may cause the behavior seen in the observed light curves. The inner disk rotates more rapidly than the magnetosphere and it leads to non-stationary behavior of matter in the inner disk and correponding non-stationary behavior of the hot spot. The hot spots are predominantly crescent-shaped but may become asymmetric (see Figure 4). Also resulting from the difference in the disk and magnetosphere rotation, the non-stationarity leads to variation of the density distribution in the hot spot that especially influences the densest parts of the spot, which may occasionally disappear. This may explain why the UV emission (associated with the high-density region of the hot spot) disappears in Figure 1.

**Data Availability**  The raw data for the *Swift*, *HST*, *TESS*, and LCOGT data are publicly available from their respective archives. The LDT raw data, CHIRON raw data, and reduced *Swift*, *HST*, *TESS*, and LCOGT data can be requested from C.C.E. Data for the accretion shock and accretion flow modeling can be requested from C.C.E. Data relevant to the 3D MHD numerical simulations can be requested from M.M.R.

**Acknowledgements**  C.C.E. acknowledges support from HST grant GO-16010, NASA grant 80NSSC19K1712, and NSF Career grant AST-1455042. M.M.R. acknowledges the NSF grant AST-20009820. N.C. acknowledges support from NASA grant NNX17AE57G.

**Author Contributions**  C.C.E. planned the observing campaign, analyzed and interpreted the data, and wrote the manuscript. C.E.R. reduced the *HST*, *TESS*, and CHIRON data, provided related text, and did the *HST* data modeling. M.M.R. did the 3D MHD simulations, provided related text, and aided in the interpretation of the data. T.T. did the CHIRON data modeling and provided related text. J.W. reduced the LCOGT and LDT data and did the timing analysis. N.C. aided in the interpretation of the data. M.R. reduced the *Swift* data. J.M. aided in the interpretation of the CHIRON data. All authors participated in the discussion of results and revision of the manuscript.

**Competing Interests**  The authors declare no competing interests.

**Correspondence**  Correspondence and requests for materials should be addressed to C. C. E.

**Extended Data Figure 1**

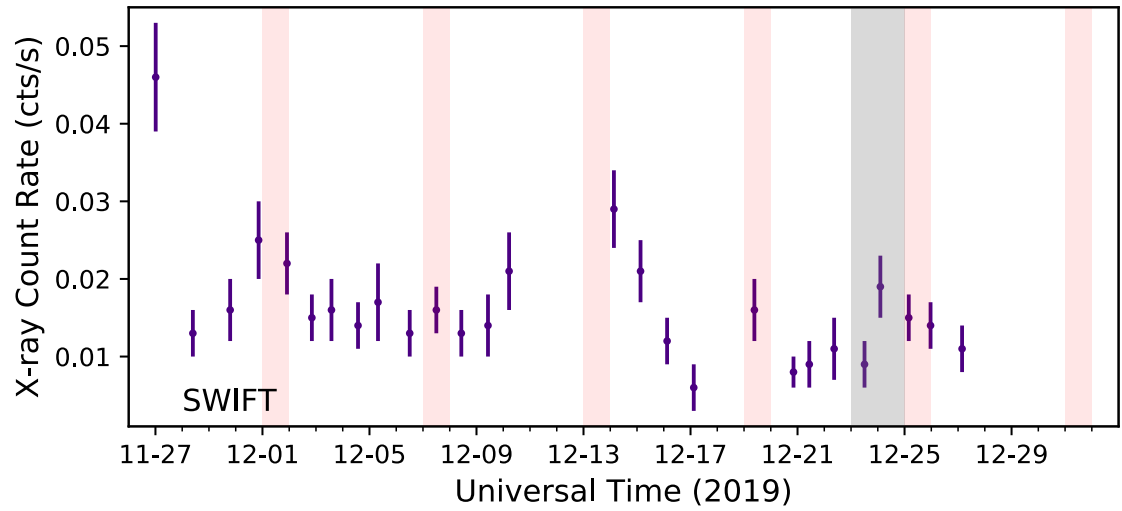

X-ray count rates of GM Aur observed by *Swift* between 2019 Nov 26 to 2020 Jan 01. To facilitate comparison, we use the same time range and color bars as those in Figure 1 for the *Swift* NUV data.

**Extended Data Figure 2**

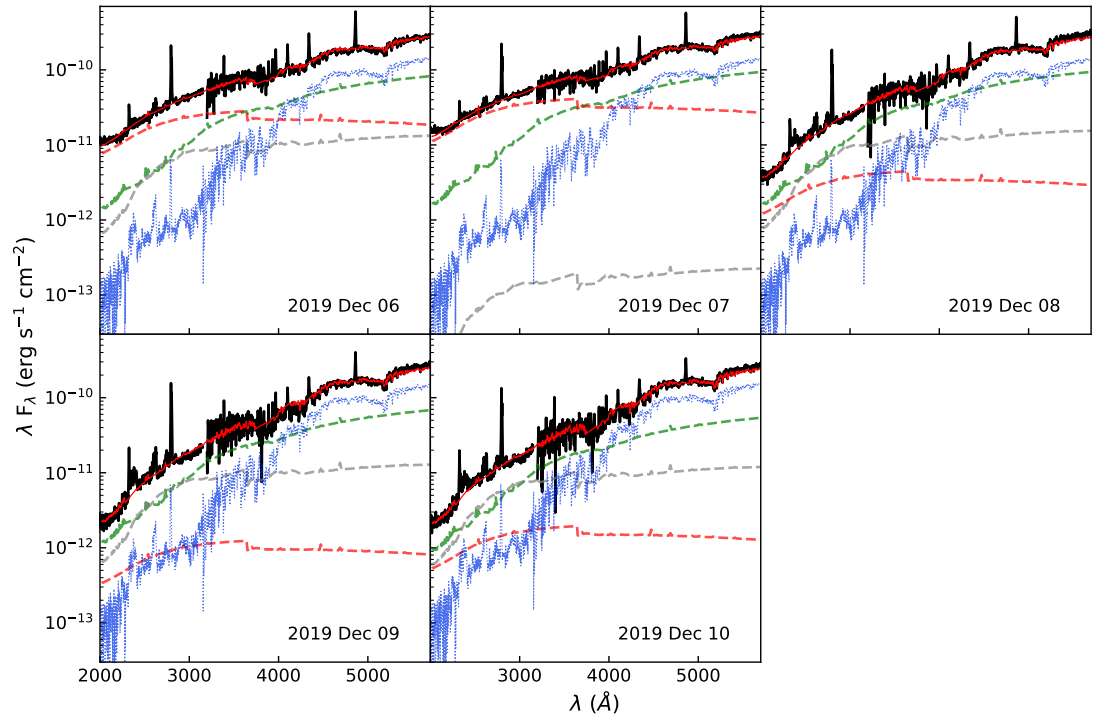

Accretion shock model fits (red solid line) to HST NUV-NIR spectra (black solid line). The total model consists of a non-accreting template star (blue dotted line) and three components with energy fluxes, $\mathcal{F}$, of $1\times10^{10}$ (green dashed line), $1\times10^{11}$ (gray dashed line), and $1\times10^{12}$ (red dashed line) erg s$^{-1}$ cm$^{-2}$. Best-fit model parameters are listed in Extended Data Table 2.

**Extended Data Figure 3**

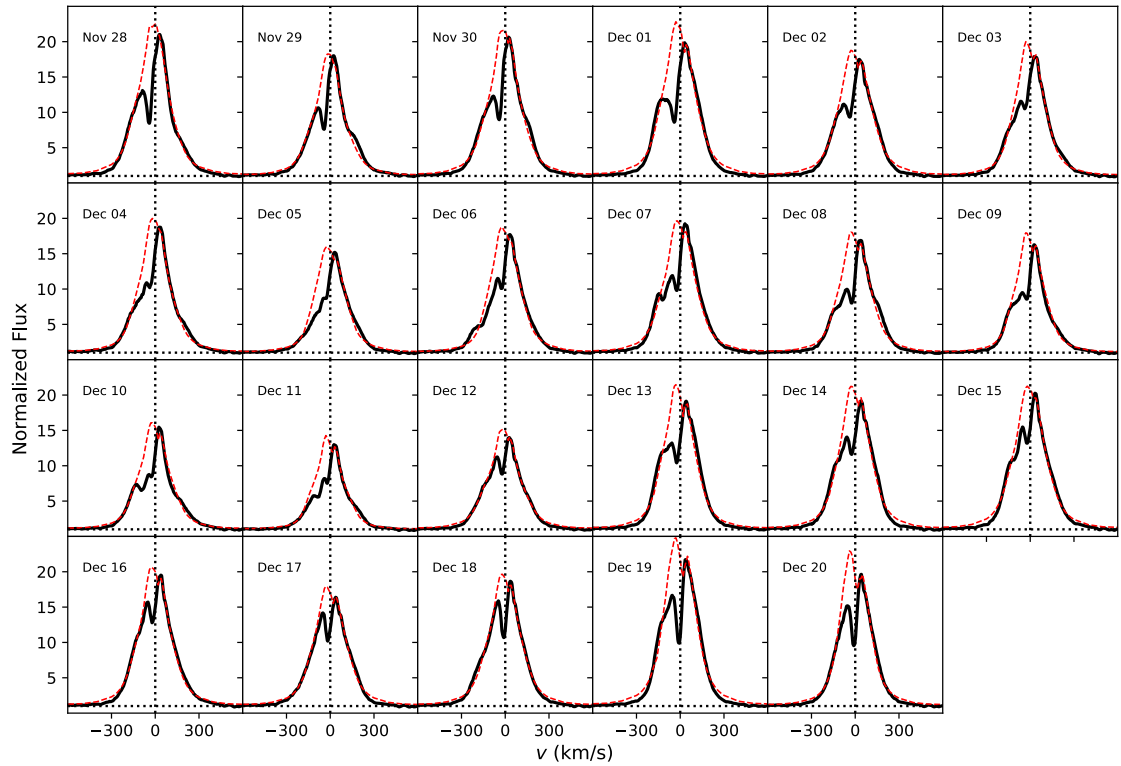

Accretion flow model fits (red broken lines) to Hα line profiles (black solid lines) measured with CHIRON. Best-fit model parameters are listed in Extended Data Table 3. There is evidence for a wind which is absorbing part of the Hα line profile on the blue side.

**Extended Data Table 1: Swift Observations and Fluxes**

| Obs. ID | Exp. Time (s) | Start Time (UT) | X-ray Count Rates (cts/s) | NUV Flux (mJy) |
|---|---|---|---|---|
| 00034249045 | 988.92704 | 2019-11-27T00:12:19 | 0.046±0.007 | 0.547±0.016 |
| 00034249046 | 1308.57631 | 2019-11-28T09:33:00 | 0.013±0.003 | 0.890±0.022 |
| 00034249047 | 1278.60907 | 2019-11-29T19:00:29 | 0.016±0.004 | 1.010±0.025 |
| 00034249048 | 1138.76270 | 2019-11-30T20:32:49 | 0.025±0.005 | 1.981±0.046 |
| 00034249049 | 1545.81653 | 2019-12-01T21:58:58 | 0.022±0.004 | 1.601±0.037 |
| 00034249050 | 1530.83276 | 2019-12-02T20:18:17 | 0.015±0.003 | 0.825±0.020 |
| 00034249051 | 1086.31981 | 2019-12-03T13:52:42 | 0.016±0.004 | 0.860±0.022 |
| 00034249052 | 1333.54882 | 2019-12-04T13:42:35 | 0.014±0.003 | 0.983±0.024 |
| 00034249053 | 659.28785 | 2019-12-05T07:37:49 | 0.017±0.005 | 0.717±0.022 |
| 00034249054 | 1428.44487 | 2019-12-06T12:04:59 | 0.013±0.003 | 1.760±0.040 |
| 00034249055 | 1443.42813 | 2019-12-07T11:57:44 | 0.016±0.003 | 1.882±0.043 |
| 00034249056 | 1238.65342 | 2019-12-08T10:22:09 | 0.013±0.003 | 0.818±0.021 |
| 00034249057 | 1146.25422 | 2019-12-09T10:17:41 | 0.014±0.004 | 0.545±0.015 |
| 00034249058 | 998.91566 | 2019-12-10T05:14:08 | 0.021±0.005 | 0.441±0.014 |
| 00034249060 | 1395.99066 | 2019-12-14T03:16:01 | 0.029±0.005 | 0.837±0.021 |
| 00034249061 | 1493.38366 | 2019-12-15T03:08:33 | 0.021±0.004 | 0.683±0.018 |
| 00034249062 | 1475.90302 | 2019-12-16T02:58:31 | 0.012±0.003 | 0.877±0.022 |
| 00034249063 | 941.47858 | 2019-12-17T02:53:06 | 0.006±0.003 | 0.587±0.017 |
| 00034249065 | 1216.17749 | 2019-12-19T09:16:32 | 0.016±0.004 | 1.890±0.044 |
| 00034249066 | 1328.55418 | 2019-12-20T20:21:19 | 0.008±0.002 | 0.735±0.019 |
| 00034249067 | 1313.57088 | 2019-12-21T10:31:53 | 0.009±0.003 | 0.813±0.021 |
| 00034249068 | 906.52676 | 2019-12-22T08:48:54 | 0.011±0.004 | 0.857±0.023 |
| 00034249069 | 1470.89856 | 2019-12-23T12:04:16 | 0.009±0.003 | 0.492±0.013 |
| 00034249070 | 1073.84346 | 2019-12-24T02:16:51 | 0.019±0.004 | 0.472±0.014 |
| 00034249072 | 1440.94136 | 2019-12-25T03:45:28 | 0.015±0.003 | 1.000±0.024 |
| 00034249073 | 1253.64681 | 2019-12-25T23:18:34 | 0.014±0.003 | 0.641±0.017 |
| 00034249074 | 1398.48779 | 2019-12-27T03:35:31 | 0.011±0.003 | 0.581±0.016 |

**Extended Data Table 2: Accretion Shock Model Parameters**

| Date | $f(1\times10^{10})$ | $f(1\times10^{11})$ | $f(1\times10^{12})$ | $\mathcal{F}_{total}$ (erg s$^{-1}$ cm$^{-2}$) | $f_{total}$ | Accretion rate ($10^{-8}$ M$_{sun}$/yr) |
|---|---|---|---|---|---|---|
| 12-6 | 0.157 | 0.005 | 0.001 | $3.3\times10^9$ | 0.163 | 1.2 |
| 12-7 | 0.176 | $9\times10^{-5}$ | 0.002 | $3.6\times10^9$ | 0.178 | 1.3 |
| 12-8 | 0.179 | 0.006 | $1.9\times10^{-4}$ | $2.6\times10^9$ | 0.185 | 0.94 |
| 12-9 | 0.127 | 0.005 | $5.3\times10^{-5}$ | $1.8\times10^9$ | 0.132 | 0.67 |
| 12-10 | 0.101 | 0.005 | $8.7\times10^{-5}$ | $1.6\times10^9$ | 0.106 | 0.57 |

**Extended Data Table 3: Accretion Flow Model Parameters**

| Date (2019 UT) | Accretion Rate ($10^{-8}$ $M_{sun}$/yr) | $R_{mag}$ ($R_{star}$) | $W_r$ ($R_{star}$) | $T_{max}$ (K) | $i$ (°) |
|---|---|---|---|---|---|
| 11-28 | 1.1 | 4.6 | 0.3 | 9060 | 65.6 |
| 11-29 | 1.0 | 3.8 | 0.2 | 8960 | 58.4 |
| 11-30 | 1.2 | 4.2 | 0.2 | 9120 | 60.4 |
| 12-01 | 1.6 | 4.1 | 0.3 | 8880 | 53.4 |
| 12-02 | 1.3 | 3.6 | 0.2 | 9010 | 48.1 |
| 12-03 | 1.3 | 3.7 | 0.2 | 8850 | 49.8 |
| 12-04 | 1.1 | 3.8 | 0.2 | 8910 | 52.6 |
| 12-05 | 1.2 | 3.4 | 0.2 | 8600 | 45.6 |
| 12-06 | 1.4 | 3.6 | 0.2 | 8680 | 47.1 |
| 12-07 | 1.4 | 3.7 | 0.2 | 8800 | 49.4 |
| 12-08 | 1.2 | 3.5 | 0.2 | 8870 | 48.2 |
| 12-09 | 1.1 | 3.9 | 0.2 | 8500 | 53.9 |
| 12-10 | 1.0 | 3.8 | 0.2 | 8750 | 61.1 |
| 12-11 | 1.2 | 3.5 | 0.2 | 8270 | 51.7 |
| 12-12 | 1.2 | 3.6 | 0.3 | 8880 | 60.2 |
| 12-13 | 1.5 | 4.0 | 0.4 | 8800 | 49.9 |
| 12-14 | 1.6 | 3.9 | 0.3 | 8760 | 48.3 |
| 12-15 | 1.3 | 4.0 | 0.2 | 8910 | 53.7 |
| 12-16 | 1.2 | 3.9 | 0.3 | 9010 | 52.2 |
| 12-17 | 1.4 | 3.5 | 0.3 | 8920 | 48.0 |
| 12-18 | 1.6 | 3.9 | 0.3 | 8700 | 53.5 |
| 12-19 | 1.9 | 4.2 | 0.5 | 8630 | 45.1 |
| 12-20 | 1.7 | 3.9 | 0.3 | 8490 | 43.6 |